\documentclass[aps,prb,superscriptaddress]{revtex4}
\usepackage{graphicx}

\def\sig{{\mbox{\boldmath{$\sigma$}}}}
\begin{document}

\title{Is telegraph noise a good model for the environment of mesoscopic systems? 
}

\author{A. Aharony}
\email{aaharonyaa@gmail.com}
\affiliation{Raymond and Beverly Sackler School of Physics and Astronomy, Tel Aviv University, Tel Aviv 69978, Israel}
\affiliation{Physics Department, Ben Gurion University, Beer Sheva 84105, Israel}

\author{O. Entin-Wohlman}
\email{oraentin@bgu.ac.il}
\affiliation{Raymond and Beverly Sackler School of Physics and Astronomy, Tel Aviv University, Tel Aviv 69978, Israel}
\affiliation{Physics Department, Ben Gurion University, Beer Sheva 84105, Israel}

\author{D. Chowdhury}
\email{debashreephys@gmail.com}
\affiliation{Physics Department, Ben Gurion University, Beer Sheva 84105, Israel}

\author{S. Dattagupta}
\affiliation{
Bose Institute, Kolkata 700054, India}
\date{\today}

\maketitle

Some papers represent the environment of a mesosopic system (e.g. a qubit in a quantum computer or a quantum junction) by a neighboring fluctuator, which generates a fluctuating electric field -- a telegraph noise (TN) -- on the electrons in the system. An example is  a two-level system, that randomly fluctuates between two states with Boltzmann weights determined by an effective temperature.  To consider whether this description is physically reasonable, we study it in the simplest example of  a quantum dot which is coupled to two electronic reservoirs and to a single fluctuator. Averaging over the histories of the  TN yields an inflow of energy flux from the fluctuator into the electronic reservoirs, which persists even when the fluctuator's effective temperature is equal to (or smaller than) the common reservoirs temperature. Therefore, the fuluctuator's temperature cannot represent a real environment. Since our formalism allows for any time dependent energy on the dot, we also apply it to the case of a non-random electric field which oscillates periodically in time. Averaging over a period of these oscillations yields results which are very similar to those of the TN model, including the energy flow into the electronic reservoirs. We conclude that both models may not give good representations of the true environment.


\section{Introduction}
\label{intro}

 This paper is devoted to the memory of Pierre Hohenberg (PCH). PCH visited Tel Aviv University in 1971, and gave a talk about the renormalization group. This talk inspired one of us (AA), then a graduate student, to move into this exciting field as a postdoc. For this, AA remained grateful to PCH ever after. After postdocing elsewhere, AA was a postdoc at Bell laboratories in Murray Hill, NJ, for a few months in 1975, working with PCH and with Bert Halperin. This resulted in  several papers on universal amplitude ratios near critical points \cite{1,2,3}, and in a later review paper in the  Domb-Lebowitz series \cite{DL}. The collaboration with Pierre was really enjoyable. He had a deep understanding of physics, and was always happy to share it. He was also a good friend, and he will be missed. During the years, both AA and SD met PCH many times, mainly at the statistical physics meetings at Yeshiva and at Rutgers, and discussions with him were always illuminating. In particular, the authors of Ref. \cite{DL} met again at Rutgers in 2003, see Fig. 1. Sadly, Vladimir Privman also passed away earlier this year.

 \begin{figure*}
 \center\includegraphics[width=6cm]{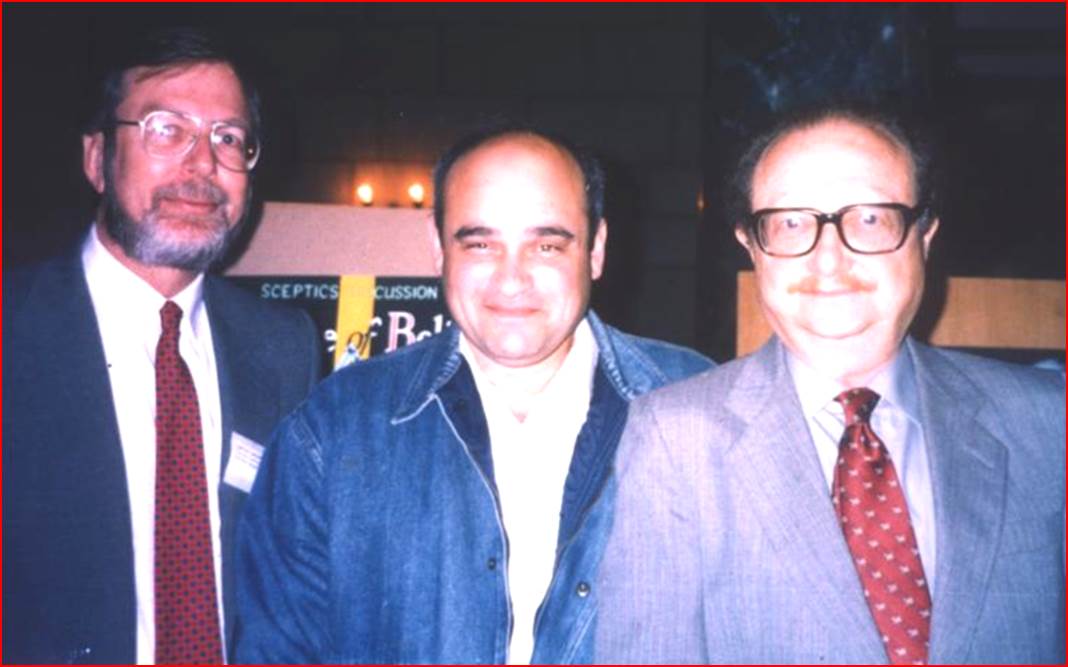}
\caption{Aharony, Privman and Hohenberg at Rutgers in 2003.}
\label{fig1}
\end{figure*}

 AA's collaboration with PCH was on critical phenomena. However, PCH had many other interests in statistical physics and in quantum mechanics. The present paper combines these two topics, in the context of mesoscopic physics. The paper raises some critical questions, which PCH would have probably enjoyed.

 Mesoscopic quantum devices usually use quantum interference, and therefore they require full coherence of their quantum states. This is particularly important in quantum computers, where the quantum information is stored in coherent states of qubits. Before using such devices for practical applications, one must overcome the dephasing and decoherence of such devices, caused by their interaction with the surrounding environment.

 A commonly used simple model for the environment concerns a single degree of freedom,
 called an "elementary fluctuator," which fluctuates
between two states. In the telegraph noise model, the fluctuator jumps randomly between these two states, independent of its earlier history \cite{SD}. The coupling between the fluctuator and the quantum system generates a potential on the system, which  then
fluctuates randomly between two values. The literature contains many possible sources for such jumps, see e.g. Refs. \cite{itakura,bergli,we2010}.
For example, TN can result from the (almost unavoidable)  presence  of defects with internal degrees of freedom, that have two (or more) metastable configurations and can switch between them due to their interaction with a thermal bath (of their own) \cite{Galperin1994}.

The TN model suffers from two important drawbacks. First, it is classical, and one does not consider its internal dynamics and its coupling to the system or to its heat bath \cite{abel,pramana,wold}. Second, the model does not consider the back action from the system onto the fluctuator. It is usually not easy to justify the neglect of the back
actions in the above  example. However, as argued by
Galperin {\it et al.} \cite{25},
 back action may be ignored when the dynamics
of the fluctuating background charge or the two-level
system is governed by its coupling to a thermalizing heat
bath, which is much stronger than its coupling to the qubit.
The TN model is also justified in the limit of a
very high temperature of this heat bath \cite{abel}. However, the problems raised below may be related to these two drawbacks.

\begin{figure}[htp]
\center\includegraphics[width=6cm]{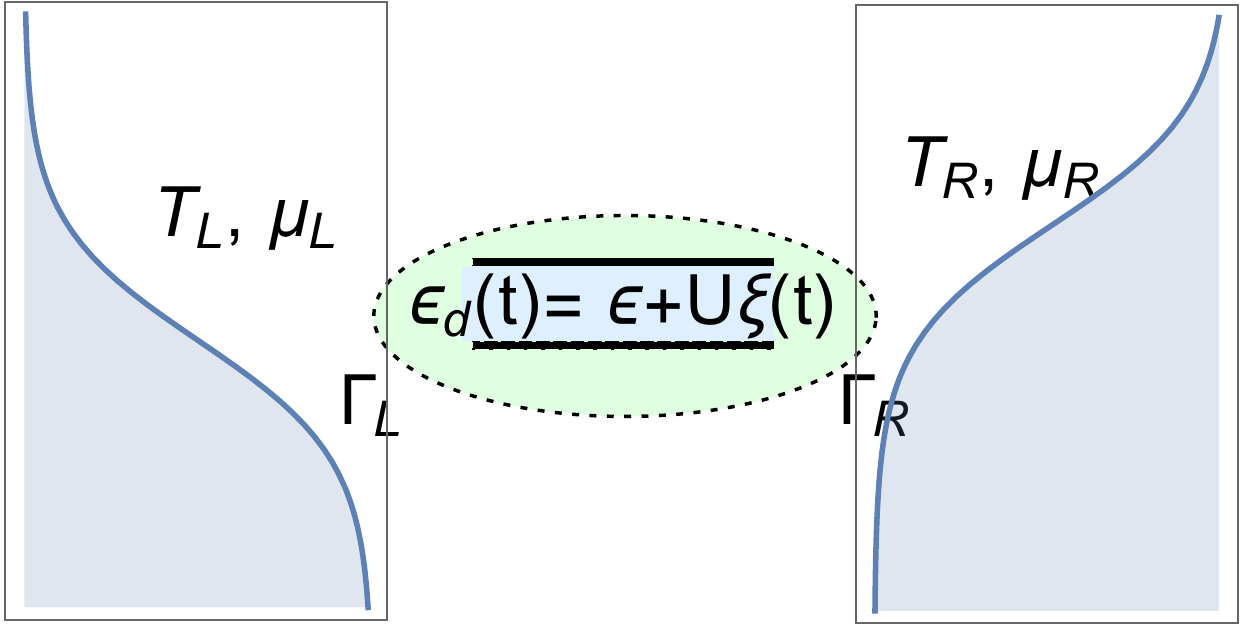}
\caption{Schematic picture of the model junction: a localized electronic level is coupled to two electronic reservoirs, held at  chemical potentials $\mu_{L}$ and $\mu_{R}$, and at  temperatures $T_{L}$ and $T_{R}$, respectively. An electron residing on this level is subjected to a stochastic electric field. This is imitated by a stochastic time dependence of the level energy,  $\epsilon_{d}(t)=\epsilon+U\xi(t)$, where $\xi(t)$ `jumps' between the values $+1$ and $-1$. The coupling with the reservoirs causes the level $\epsilon$ to become a resonance, of width $\Gamma=\Gamma_{L}+\Gamma_{R}$, where $\Gamma_{L},\Gamma_{R}$ are the partial widths. }
\label{Fig1}
\end{figure}

Naively, one might expect that when the fluctuator is coupled to a mesoscopic system, which is also coupled to other heat reservoirs, then the combined system would reach thermal equilibrium when all the heat reservoirs have the same temperature. Below we show a specific example, in which this is not the case. Specifically, we consider the charge and energy currents carried by spinless electrons between two reservoirs (also called leads or terminals) through a junction, which consists of a single two-level quantum dot, see Fig. \ref{Fig1}.  The electronic energy level on the dot fluctuates between two values,
\begin{equation}
\epsilon^{}_{d}(t)=\epsilon +U\xi(t)=\epsilon^{}_0+U[\xi(t)-\overline{\xi}]\ ,
\label{ed}
\end{equation}
where $\xi(t)$ jumps randomly between the two values $\xi=\pm 1$, $\overline{\xi}$ is its average and $\epsilon^{}_0=\epsilon+U\overline{\xi}$. Below we set $\epsilon^{}_0$ to be the zero of energies.
The fluctuating term is
due to  a random TN which is  produced by a single neighboring electronic defect. The currents are averaged over all the possible histories of the telegraph process. The average charge current, which was also calculated in Refs. \cite{Galperin1994,we2016}, shows a double peak (or a single broad peak) as function of the bias voltage between the electronic reservoirs. This result can be intuitively understood, since the electrons scattered by the quantum dot cross it when its energy level has one of the two possible values.

The situation becomes less intuitive when one considers the average heat current \cite{DC}. Surprisingly,
 it is found that the defect supplies  energy to the electronic reservoirs, which is distributed unequally between them: the stronger is the coupling of the reservoir with the junction, the more energy it  gains. Thus the noisy environment can lead to a temperature gradient across an un-biased junction. As we show below, this energy flux from the fluctuator always flows from the fluctuator into the electronic reservoirs, even when the temperature associated with the fluctuator's thermal bath is lower than the temperatures of these reservoirs. This surprising result raises questions concerning the meaning of the fluctuator's temperature, and therefore also about the physical utility of the TN model: Since the model Hamiltonian does not include the interaction between the fluctuator and the dot, nor the back action from the electrons to the fluctuator, the system is not in equilibrium and there is no meaning to a comparison of $T$ with the temperatures of the electronic reservoirs.

In some sense, the fluctuating energy level on the quantum dot can be interpreted as resulting from a random fluctuating electric potential which acts on the electrons when they visit the dot. In order to check the importance of the randomness of the fluctuating field, we repeat the same analysis for a non-random oscillating electric field, in which the energy level of the electrons on the dot has the form
\begin{equation}
\epsilon^{}_d(t)=\epsilon+U\cos(\Omega t)\ .
\label{edosc}
\end{equation}
This model has been treated in several earlier papers \cite{liliana}.
Below we treat this model using exactly the same tools as for the TN case. The only difference is the method of averaging: instead of averaging over histories we now average over one period of the oscillation. A detailed comparison between the two models shows many similarities. Both models exhibit additional peaks in the particle transmission, and both models yield an energy flow from the time-dependent field into the electronic reservoirs. Thus, it seems that the randomness of the TN model is not essential, and therefore, again, it is not clear if this model is a good representative of the environment.

Section \ref{HHH} presents a  Keldysh derivation of the particle and energy currents, for a general time-dependent energy level on the dot, $\epsilon^{}_d(t)$.
After introducing the Hamiltonian for the model of Fig. \ref{Fig1},  the various particle and energy currents are expressed in terms of time-dependent Green's functions. The detailed Dyson equations for these Green's functions are solved  in Appendix A. We then apply these general results  to our two models.  Section \ref{TN}
describes the TN model, for which the observable physical properties jump randomly between two values, while Sec. \ref{EE}  describes the results for a periodically oscillating dot energy. In both cases we also calculate the averages over these time variations: the TN is averaged over all possible histories, and the oscillating observables are averaged over a period of the oscillations.
 The physical consequences are then discussed in Sec. \ref{dis}. Although much of the information included in Secs. \ref{HHH}-\ref{TN} was contained in Ref. \cite{DC}, it is presented here in different ways, with much more detail (especially in Sec. \ref{TN}). We also present more results. Our results for the oscillating field were implicitly contained in Ref. \cite{liliana}, but here we calculate them for the special case of the wide band approximation, which yields simple analytic expressions, allowing more physical insight.

\section{Hamiltonian and currents}\label{HHH}

The total Hamiltonian of the system and the reservoirs in Fig. \ref{Fig1} is
\begin{equation}
{\cal H}={\cal H}^{}_{\rm leads}+{\cal H}^{}_{\rm sys}(t)+{\cal H}^{}_{\rm tun}\ .
\label{genH}
\end{equation}
The (time dependent) Hamiltonian of the system is
\begin{equation}
{\cal H}^{}_{\rm sys}(t)=\epsilon^{}_{d}(t)d^{\dagger}_{}d\  ,
\label{Hsys}
\end{equation}
where $d$ ($d^{\dagger}$) annihilates (creates) an electron on the level,  whose energy depends on time. This time dependence need not be specified in this section.   The dot is coupled to two electronic reservoirs of spinless electrons  by tunneling amplitudes $V_{\bf k}$ and $V_{\bf p}$,
\begin{equation}
{\cal H}^{}_{\rm tun}={\cal H}^{}_{\rm tun,L}+{\cal H}^{}_{\rm tun,R}=\sum_{\bf k}(V^{}_{\bf k}c^{\dagger}_{\bf k}d+{\rm Hc})+\sum_{\bf p}(V^{}_{\bf p}c^{\dagger}_{\bf p}d+{\rm Hc})\ ,
\label{Htun}
\end{equation}
where $c^{}_{{\bf k}({\bf p})}$  ($c^{\dagger}_{{\bf k}({\bf p})}$)  are the annihilation (creation) operators for the electrons with wave vectors ${\bf k}({\bf p})$  in the leads.
The leads are modeled as free electron gases,
\begin{equation}
{\cal H}^{}_{\rm leads}={\cal H}^{}_{\rm L}+{\cal H}^{}_{\rm R}=\sum_{\bf k}\epsilon^{}_{k}c^{\dagger}_{\bf k}c^{}_{\bf k}+
\sum_{\bf p}\epsilon^{}_{p}c^{\dagger}_{\bf p}c^{}_{\bf p}\ .
\label{Hleads}
\end{equation}

The  particle flux into the left lead, i.e., the rate of change of the number of particles there, $N^{}_L=\sum_{\bf k}c^{\dagger}_{\bf k}c^{}_{\bf k}$, is
\begin{eqnarray}
&I^{}_{L}(t)=\langle \frac{d}{dt}N^{}_L\rangle=i\langle[{\cal H},N^{}_L]\rangle
&=\sum_{\bf k}[V^{\ast}_{\bf k}G^{<}_{{\bf k}d}(t,t)-V^{}_{\bf k}G^{<}_{d{\bf k}}(t,t)]\ .
\label{IL}
\end{eqnarray}
 Here $G^{<}_{ab}(t,t')=i\langle b^{\dagger}_{}(t')a^{}_{}(t)\rangle $ is the Keldysh lesser Green's function; the angular brackets indicate the quantum average.
 From now on we assume that the tunneling amplitudes are determined by the wave vectors corresponding to the energy of the tunneling electrons. In other words, $V^{}_{\bf k}$ and $V^{}_{\bf p}$ are independent of the wave vectors, and are replaced by $V^{}_L$ and $V^{}_R$.
 The current into the right lead, $I^{}_R$, is defined similarly,  with $L\Rightarrow R$ and ${\bf k}\Rightarrow {\bf p}$.

The electronic occupation on the dot is
\begin{equation}
Q^{}_{d}(t)=\langle d^\dagger_{} d^{}_{}\rangle=-iG^{<}_{dd}(t,t)\ .
\label{Q}
\end{equation}
Since the total number of electrons is conserved, $[{\cal H},N^{}_L+N^{}_R+d^\dagger_{}d^{}_{}]=0$, it follows that the  flux of particles into the dot  is  compensated by the sum of the two fluxes into the leads,
\begin{equation}
I^{}_{d}(t)=\frac{d}{dt}Q^{}_d(t)=-[I^{}_{L}(t)+I^{}_{R}(t)]\ ,
\end{equation}
that is, particle number in the junction is conserved.

We next turn to the  energy fluxes that flow in the time domain. Each term ${\cal H}^{}_n$ in the Hamiltonian generates an energy flux,
\begin{equation}
I^E_n(t)=\langle \frac{d}{dt}{\cal H}^{}_n\rangle=\langle[{\cal H},{\cal H}^{}_n]+\frac{\partial}{\partial t}{\cal H}^{}_n\rangle\ .
\label{IEn}
\end{equation}
Since only ${\cal H}^{}_{\rm sys}$ depends explicitly on time, we have
\begin{equation}
\sum_n I^{E}_n(t)=\langle[{\cal H},{\cal H}]+\frac{\partial}{\partial t}{\cal H}^{}_{\rm sys}\rangle=\langle \frac{\partial}{\partial t}{\cal H}^{}_{\rm sys}\rangle\ ,
\label{sumI}
\end{equation}
The last term contributes only to the energy current into the dot,
\begin{equation}
I^{E}_{d}(t)=
 \frac{d}{dt}\big[\epsilon^{}_{d}(t)Q^{}_d(t)\big]\nonumber\\
=\frac{d\epsilon^{}_{d}(t)}{dt}Q^{}_d(t)+\epsilon^{}_{d}(t)I^{}_{d}(t)\ .
\label{IDE}
\end{equation}
The first term on the right hand-side of Eq. (\ref{IDE}) results from the explicit time-dependence of the localized energy, i.e., it is due to time-dependent electric potential acting on the dot. {\it This term expresses the power supplied to the system by the field}. We  denote this power by
\begin{equation}
P^{}_{d}(t)=Q^{}_{d}(t)\frac{d\epsilon^{}_{d}(t)}{dt}\ .
\label{P}
\end{equation}
Equation (\ref{sumI}) thus becomes \cite{liliana}
\begin{equation}
I^{E}_{L}(t)+I^{E}_{R}(t)+I^{E}_{{\rm tun},L}(t)+I^{E}_{{\rm tun},R}(t)+I^{E}_{d}(t)=P^{}_{d}(t)\ ,
\end{equation}
which expresses the energy conservation in the  junction.

Specifically, the energy current into the left reservoir is
\begin{equation}
I^{E}_{L}(t)
=\Big\langle \frac{d}{dt}\sum_{\bf k}\epsilon^{}_{k}c^{\dagger}_{\bf k}c^{}_{\bf k}\Big\rangle
=\sum_{\bf k}\Big (\epsilon^{}_{k}[V^{\ast}_{\bf k}G^{<}_{{\bf k}d}(t,t)-V^{}_{\bf k}G^{<}_{d{\bf k}}(t,t)]\Big )\ .
\label{ILE}
\end{equation}
The analogous energy flux associated with the right reservoir is derived from Eq. (\ref{ILE})
by interchanging $L\Leftrightarrow R$ and ${\bf k}\Leftrightarrow {\bf p}$.

Unlike the particle currents, the time-dependent energy fluxes also include the temporal variation of the (left and right) tunneling Hamiltonians, Eq. (\ref{Htun}),
\begin{eqnarray}
&I^{E}_{{\rm tun}, L}(t)= \langle \frac{d}{dt}\sum_{\bf k}(V^{}_{\bf k}c^{\dagger}_{\bf k}d+{\rm Hc})\rangle
=\epsilon^{}_{d}(t)I^{}_{L}(t)-I^{E}_{L}(t)\nonumber\\
&+\sum_{{\bf k},{\bf p}}[V^{\ast}_{\bf k}V^{}_{\bf p}G^{<}_{{\bf k}{\bf p}}(t,t)-V^{}_{\bf k}V^{\ast}_{\bf p}G^{<}_{{\bf p}{\bf k}}(t,t)]\ ,
\label{ITLE}
\end{eqnarray}
(with an analogous expression for $I^{E}_{{\rm tun},R}$). However, as we show below, the temporal averages of these contributions vanish.

Appendix A presents all the currents in terms of the three Green's functions on the dot, $G^{}_{dd}(t,t')$, within the wide band approximation.
Specifically, the retarded (advanced) Green's function on the dot is
\begin{equation}
G^{r(a)}_{dd}(t,t')=\mp i \Theta (\pm t\mp t')
e^{-i\int_{t'}^{t}dt^{}_{1}\epsilon^{}_{d}(t^{}_{1})\mp \Gamma (t-t')}
\ ,
\label{GDRA}
\end{equation}
where
\begin{equation}
\Gamma=\Gamma^{}_L+\Gamma^{}_R\ ,\ \ \ \Gamma^{}_{L(R)}=\pi {\cal N}^{}_{L(R)}|V^{}_{L(R)}|^2
\end{equation}
(${\cal N}^{}_{L(R)}$ is the density of states in reservoir $L(R)$).
Both the retarded and the advanced dot Green's functions depend on  the time-dependent energies $\epsilon^{}_d(t)$ only through the function
\begin{equation}
X(t,t')=e^{i\int_{t'}^t dt''\epsilon_d(t'')}\ .
\label{XX}
\end{equation}
For the two types of averaging considered in the next two sections,  the average of $X(t,t')$  depends only on the time difference, $(t-t')$, and therefore we denote it by $\overline{X(t,t')}\equiv \overline{X}(t-t')$.
 The averages of Eqs. (\ref{GDRA}) thus become
 \begin{eqnarray}
 \overline{G^a_{dd}(t-\tau,t)}=i\Theta(\tau)\overline{X}(\tau)e^{-\Gamma\tau}\ ,
 \ \ \ \ \overline{G^r_{dd}(t,t-\tau)}=[\overline{G^a_{dd}(t-\tau,t)}]^\ast\ ,
 \end{eqnarray}
 independent of $t$.
 Below we shall need the Fourier transforms of these functions,
 \begin{eqnarray}
&\overline{G^{a}_{dd}(\omega)}=\int d\tau e^{-i\omega\tau}\overline{G^{a}_{dd}(t-\tau,t)}=i\int_0^\infty d\tau e^{-(i\omega+\Gamma)\tau}\overline{X}(\tau)\ ,\nonumber\\
&\overline{G^{r}_{dd}(\omega)}=\int d\tau e^{i\omega\tau}\overline{G^{r}_{dd}(t,t-\tau)}=[\overline{G^a_{dd}(\omega)}]^\ast\ .
\end{eqnarray}
 Defining the Laplace transform of a function $F(t)$ by
 \begin{equation}
\widetilde{F}(s)\equiv\int_0^\infty d\tau e^{-s\tau}F(\tau)\ ,
\label{LT}
\end{equation}
 we identify
 \begin{equation}
 \overline{G^{a}_{dd}(\omega)}=i\widetilde{\overline{X}}(\Gamma+i\omega)\ .
\label{LTGa}
\end{equation}

Similarly, Appendix A yields the lesser Green's function on the dot,
\begin{equation}
G^<_{dd}(t,t)=\int\frac{d\omega}{2\pi}\Sigma^<(\omega)[K(t,\omega)+cc]\ ,
\label{GDless}
\end{equation}
where
\begin{equation}
K(t,\omega)\equiv \int_{-\infty}^t dt^{}_1 e^{2\Gamma(t^{}_1-t)}\int_0^\infty d\tau e^{-(\Gamma+i\omega)\tau}X(t^{}_1,t^{}_1-\tau)\ .
\label{KKK}
\end{equation}

We now consider the average, $\overline{G^<_{dd}(t,t)}$. Since $\overline{X(t^{}_1,t^{}_1-\tau)}$ depends only on $\tau$, the integral over $\tau$ yields $\widetilde{\overline{X}}(\Gamma+i\omega)$, independent of $t^{}_1$, hence
 \begin{eqnarray}
& \overline{G^<_{dd}(t,t)}=i\overline{Q^{}_d}=\int\frac{d\omega}{2\pi}\Sigma^<(\omega){\rm Re}\big[\widetilde{\overline{X}}(\Gamma+i\omega)\big]\Gamma\nonumber\\
& =2i\int\frac{d\omega}{2\pi}[\Gamma^{}_Lf^{}_L(\omega)+\Gamma^{}_Rf^{}_R(\omega)]{\rm Im}[\overline{G^a_{dd}(\omega)}]/\Gamma\ .
\label{QQdd}
 \end{eqnarray}
The average occupation on the dot,
\begin{equation}
\overline{Q^{}_d}=-i\overline{G^<_{dd}(t,t)}=2\int\frac{d\omega}{2\pi}[\Gamma^{}_Lf^{}_L(\omega)+\Gamma^{}_Rf^{}_R(\omega)]{\rm Re}[\widetilde{\overline{X}}(\Gamma+i\omega)]/\Gamma\ ,
\label{QQd}
\end{equation}
is thus independent of $t$.

Substituting all these Green's function into Eq. (\ref{IL0}),  the average particle current into the left reservoir becomes
\begin{equation}
\overline{I^{}_L}=\int\frac{d\omega}{2\pi}[f^{}_R(\omega)-f^{}_L(\omega)]{\cal T}(\omega)\ ,
\label{IILL}
\end{equation}
with the average transmission
\begin{equation}
{\cal T}(\omega)=\frac{4\Gamma^{}_L\Gamma^{}_R}{\Gamma}{\rm Im}[\overline{G^{a}_{dd}(\omega)}]=\frac{4\Gamma^{}_L\Gamma^{}_R}{\Gamma}{\rm Re}[\widetilde{\overline{X}}(\Gamma+i\omega)]\ .
\label{TTT}
\end{equation}
Without the time-dependent part in ${\cal H}^{}_{\rm sys}$, ${\cal T}(\omega)$ shows a simple single Breit-Wigner peak. As we show below, the time-dependent parts split this peak. In the TN case, the two new peaks are related to the two shifted levels. In the oscillating field case there appear peaks at energies which are shifted by integer multiples of the oscillation frequency.

 Appendix A also presents the derivation of the various energy currents, specifically
\begin{eqnarray}
&\overline{I^{E}_{L}}=2i\Gamma^{}_{L}\Big (
\int\frac{d\omega}{2\pi}\omega f^{}_{L}(\omega)[\overline{G^{a}_{dd}}(\omega)-\overline{G^{r}_{dd}}(\omega)]
-\overline{\epsilon^{}_{d}(t)
G^{<}_{dd}(t,t)}\Big )
\nonumber\\
&-i\Gamma^{}_{L}\int\frac{d\omega}{2\pi}\Sigma^{<}_{}(\omega)[\overline{G^{a}_{dd}}(\omega)+\overline{G^{r}_{dd}}(\omega)]\ .
\label{ILEWt}
\end{eqnarray}
For the two special types of averages, presented in the next two sections, one can use the explicit averages for the various terms, and show that
\begin{equation}
\overline{I^{E}_{L}}=\int\frac{d\omega}{2\pi}[f^{}_{R}(\omega)-f^{}_{L}(\omega)]\omega{\cal T}(\omega)+\frac{\Gamma^{}_{L}}{\Gamma}\overline{P^{}_{d}}
\ .
\label{IELW}
\end{equation}
  The first term, which vanishes unless the junction is biased (by a voltage and/or temperature difference), is  indeed  the usual  energy current in a two-terminal electronic junction, with the transmission ${\cal T}(\omega)$ given in Eq. (\ref{TTT}). The second term comes from the power supplied by the source of the  fluctuating processes, Eq. (\ref{P}). The sign of the first term depends on the bias, and/or  on  the temperature difference across the junction; this means that energy flux represented by the first term  can flow out or into the left reservoir. 
As opposed,
$\overline{P_{d}}$ is found to be positive for both models, which means that energy flows into the left reservoir from the time-dependent sources.
Since
$\overline{I^{E}_{R}}$ is given by Eq. (\ref{IELW}) with $L\Leftrightarrow R$, it follows that energy flows also into the right reservoir. Thus, the telegraph  noise supplies energies to both reservoirs, with the larger portion going into the more  strongly-coupled one.
Although we do not have a general proof, we conjecture that  Eq. (\ref{IELW}) probably holds for a general time-dependent source of energy on the dot.

\section{The telegraph noise}
\label{TN}

In the TN model, the localized energy on the dot in our junction is given by Eq. (\ref{ed}).
We start with a detailed tutorial on the conditional average method to solve this model.
Beginning at an initial time $t^{}_0$, the function $\xi(t)$ jumps instantaneously between the values $+1$ and $-1$ at random instants $t^{}_0< t^{}_{1}<t^{}_{2}<t^{}_{3}<\ldots<t$.
Each history of the system involves a certain sequence of times, at which the jumps occur.
In our calculation, we average the time-dependent fluxes (at time $t$) over all histories; this procedure amounts to averaging over all the time sequences, time-dependent probabilities for jumps to (or not to) occur in given time intervals and over the two possible initial values of the random variable $\xi(t^{}_0)$. Below we take $t^{}_0\rightarrow -\infty$. The average contains the case with no jump, the case with one jump at any intermediate time between the initial time $t_{0}$ and $t$,  the case with two jumps at any intermediate times $t^{}_0<t^{}_1<t^{}_2<t$, {\it etc}.

In the simplest example, the TN is characterized by the {\it a priori} probabilities of the occurrence of $\xi=+1$ (or $-1$).
In the example of the elementary charge fluctuators, each fluctuator is assumed to be in one of two states, with average probabilities $p^{}_+$ or $p^{}_-$. For example, these states can represent a charged particle jumping between two close potential minima, generating  an electrostatic potential $+U$ or $-U$ on the electron which occupies the quantum dot. If the two states of the fluctuator have energies $\pm E_0/2$ (with $E_0>0$), then their average probabilities $p^{}_\pm$ are determined by their interaction with a separate heat bath of temperature $T$, then the probabilities are
given by the (normalized) Boltzmann factors \cite{Galperin1994},
\begin{equation}
p^{}_{\pm}(T)=\frac{\exp[\pm E_0/(2k^{}_{\rm B}T)]}{2{\rm cosh}[E_0/(2k^{}_{\rm B}T)]}\ .
\label{p}
\end{equation}
The ``TN temperature" $T$ is an {\it effective} temperature, that models the probabilities $p^{}_\pm$.  At $T=0$ (and $E^{}_0>0$), Eq. (\ref{p}) yields  $p^{}_+=1$, hence no fluctuations. The effect of the fluctuations then increases as $T$ is raised. Other models of the TN give similar results. With these probabilities, the average of $\xi(t)$ is independent of the time,
\begin{equation}
\overline{\xi}=p^{}_{+}(T)-p^{}_{-}(T)={\rm tanh}[E_0/(2k^{}_{\rm B}T)]\ .
\label{overx}
 \end{equation}
(We denote  an average over the telegraph  process  by an over bar, to distinguish it from the quantum average, which is indicated by angular brackets.) As $T\rightarrow\infty$, the average $\overline{\xi}$ tends to zero. As $T\rightarrow 0$, the fluctuator remains at its ground state and $\xi(t)=\overline{\xi}=1$.

In addition to the probabilities $p^{}_\pm$, one also needs to specify the mean rate $\gamma$ at which the instantaneous jumps occur. Assuming detailed balance,  the probability per unit time to jump from $\xi=a(=\pm 1)$ to $\xi=b(=\mp 1)$ is $W^{}_{ab}=\gamma p^{}_b$, and the probability per unit time to stay in the state $a$ is  $-W^{}_{aa}=\sum_{b\ne a}W_{ab}=\gamma p^{}_{-a}$. The total rate for any jump is $\gamma=W^{}_{1,-1}+W^{}_{-1,1}$. It is expedient to present the four possible values of $W_{ab}$ in a matrix form,
\begin{equation}
{\bf W}=\gamma\left[\begin{array}{cc}-p^{}_- &\ \ \  p^{}_- \\ \ p^{}_+ &\  -p^{}_+ \end{array}\right ]\  .
\label{WW}
\end{equation}

The conditional average probability that $\xi(t)=b$, given that $\xi(t')=a$, is the $2\times 2$ matrix ${\bf P}(t,t')$,  which solves the differential equation
 \begin{equation}
 \frac{d}{dt}{\bf P}(t,t')={\bf W P}(t,t')\ ,
 \label{dP}
 \end{equation}
 with the initial condition ${\bf P}(t',t')={\bf I}$, the $2\times 2$ unit matrix. A convenient way to solve this equation is to use its Laplace transform, Eq. (\ref{LT}). From Eq. (\ref{dP}),
 \begin{equation}
 \widetilde{\bf P}(s)=[s{\bf I}-{\bf W}]^{-1}=\frac{1}{s(s+\gamma)}\left[\begin{array}{cc}s+\gamma p^{}_+ &\ \ \ \gamma p^{}_- \\ \ \gamma p^{}_+ &\  s+\gamma p^{}_- \end{array}\right ]\ ,
 \label{LP}
 \end{equation}
 with the inverse Laplace transform
 \begin{equation}
 {\bf P}(t,t')=\left[\begin{array}{cc}p^{}_++ p^{}_-e^{-\gamma (t-t')} &\  p^{}_-[1-e^{-\gamma (t-t')}] \\ \ p^{}_+[1-e^{-\gamma (t-t')}] & \ \ p^{}_-+p^{}_+e^{-\gamma (t-t')} \end{array}\right ]\ ,
 \end{equation}
 which depends only on the difference $(t-t')>0$. This can also be written as
 \begin{equation}
 \label{PP}
 {\bf P}(t,t-\tau)\equiv {\bf P}(\tau)=e^{{\bf W}\tau}\nonumber\\
 ={\bf T}+({\bf I-T})e^{-\gamma \tau}\ ,
 \end{equation}
where
\begin{equation}
 {\bf T}\equiv {\bf I+W}/\gamma=\left[\begin{array}{cc} p^{}_+ \ \ \  p^{}_- \\  p^{}_+ \ \ \ p^{}_- \end{array}\right ]\  .
 \label{T}
 \end{equation}
Thus, ${\bf P}(\infty)={\bf T}$.
It is easy to confirm that ${\bf T}$ and $({\bf I-T})$ are projection matrices, ${\bf T}^2={\bf T}$, $({\bf I-T})^2={\bf I-T}$ and ${\bf T}({\bf I-T})=0$, and therefore
\begin{equation}
{\bf P}(t^{}_1){\bf P}(t^{}_2)={\bf P}(t^{}_1+t^{}_2)\ ,
\label{PP}
\end{equation}
 as expected.

To average over the TN histories, it is convenient to define {\it conditional averages}:  the average of the function $Z(t,t')$ under the assumption that $\xi(t')=a$ and $\xi(t)=b$ (with $t>t'$) is expected to depend only on $t-t'$, and is hence denoted by $Z(t-t')^{}_{ab}$. The average over all histories and all initial and final values of $\xi$ is \cite{Blume}
 \begin{equation}
 \label{av}
 \overline{Z(t,t')}=\sum_{a,b}p^{}_aZ(t-t')^{}_{ab}= {\rm Tr}\big[{\bf TZ}(t-t')\big]\ .
 \end{equation}
Therefore, the average is also only a function of the time difference $(t-t')$, and it is invariant under time translations.

In the above derivation we assumed that at any initial time the fluctuator is at equilibrium with its bath, and therefore the initial probabilities $p^{}_a$ do not depend on the initial time $t'$. However, as we now show, this assumption is not necessary.
In principle, we should average over the TN between the times $t^{}_0=-\infty$ and $t$. However, as seen in Sec. \ref{HHH}, some quantities [e.g. $X(t'',t')$] depend on the random telegraph variables only between intermediate times, e.g. $t'$ and $t''$, with $t^{}_0<t'<t''<t$. The conditional average of  $Z(t'',t')$, when the average is over the whole time span, should therefore be
 \begin{equation}
 \overline{Z(t'',t')}={\rm Tr}\big[{\bf TP}(t'-t^{}_0){\bf Z}(t'',t'){\bf P}(t-t'')\big]\ .
 \end{equation}
 (We multiply matrices with increasing times from left to right; one could also construct the whole theory with the opposite order of matrices). However, using the properties of the trace, Eq. (\ref{PP}) and ${\bf P}(\infty)={\bf T}$ this becomes
 \begin{eqnarray}
 &\overline{Z(t'',t')}={\rm Tr}\big[{\bf P}(t-t''){\bf TP}(t'-t^{}_0){\bf Z}(t'',t')\big]={\rm Tr}\big[{\bf P}(t-t''+t'+\infty){\bf Z}(t'',t')\big]\nonumber\\
& ={\rm Tr}\big[{\bf T}{\bf Z}(t'',t')\big]={\rm Tr}\big[{\bf TZ}(t''-t')\big]\ ,
 \end{eqnarray}
which is the same as in Eq. (\ref{av}). Thus,  the results of the averaging depend only on the time interval for which the average is calculated.
 We shall use this property to average over quantities inside integrals, see below.

As seen in the previous section, many quantities which need to be averaged contain the factor $X(t,t')$, defined in Eq. (\ref{XX}).
 Such factors and their averaging have been discussed in the literature since a long time
 \cite{Blume,anderson,Galperin1994,DC}. We now average $X(t,t')$ over all the histories of the TN.
 Using Eq. (\ref{ed}), with  $\epsilon^{}_0\equiv\epsilon+U\overline{\xi}=0$, we have
 \begin{equation}
 X(t,t')=e^{iU\int_{t'}^t d\tau[\xi(\tau)-\overline{\xi}]}\ .
 \end{equation}
Since $dX(t,t')/dt=iU[\xi(t)-\overline{\xi}]X(t,t')$, integration yields \cite{Blume}
\begin{equation}
X(t,t')= 1+iU\int_{t'}^{t}d\tau[\xi(\tau)-\overline{\xi}]X(\tau,t')\ .
\end{equation}
The conditional average of this equation is expected to depend only on $t-t'$, and it obeys the equation
\begin{equation}
X(t-t')^{}_{ab}=P(t-t')^{}_{ab}\nonumber\\
+iU\sum_{c}\int_{t'}^{t}d\tau X(\tau-t')^{}_{ac}[c-\overline{\xi}]P(t-\tau)^{}_{cb}\ .
\end{equation}
In matrix form,
\begin{equation}
{\bf X}(t-t')={\bf P}(t-t')\nonumber\\
+iU\int_{t'}^{t}d\tau {\bf X}(\tau-t')[\sig^{}_z-\overline{\xi}{\bf I}]{\bf P}(t-\tau)\ ,
\label{Xttp}
\end{equation}
where $\sig^{}_z$ is the Pauli matrix.
Equation (\ref{Xttp}) is conveniently solved by Laplace transforming it   \cite{Blume},
\begin{equation}
\widetilde{\bf X}(s)=\widetilde{\bf P}(s)+iU\widetilde{\bf X}(s)(\sig^{}_z-\overline{\xi}{\bf I})\widetilde{\bf P}(s)\ .
\end{equation}
Using Eq. (\ref{LP}),
the solution for the matrix $\widetilde{\bf X}(s)$ is
\begin{equation}
\widetilde{\bf X}(s)=[\widetilde{\bf P}(s)^{-1}-iU(\sig^{}_z-\overline{\xi}{\bf I})]^{-1}\nonumber\\
=\frac{1}{D}\left[\begin{array}{cc}s+iU(\overline{\xi}+1)+\gamma p^{}_+ & \gamma p^{}_- \\ \gamma p^{}_+ & s+iU(\overline{\xi}-1)+\gamma p^{}_- \end{array}\right ]\  ,
\label{}
\end{equation}
with the determinant
\begin{equation}
D=(s+iU\overline{\xi})(s+iU\overline{\xi}+\gamma)+U^2-iU\gamma\overline{\xi}\ .
\end{equation}
It follows from  Eq. (\ref{av})  that \cite{com2}
\begin{equation}
\overline{\widetilde{ X}}(s)=\frac{s+\gamma+2iU\overline{\xi}}{D}=\frac{1}{2u}\Big[\frac{\gamma+2iU\overline{\xi}+s^{}_+}{s-s^{}_+}-\frac{\gamma+2iU\overline{\xi}+s^{}_-}{s-s^{}_-}\Big]\ ,
\label{fX}
\end{equation}
where
\begin{equation}
s_\pm=-\gamma/2-i U \overline{\xi}\pm u,\ \ \ \ u=\sqrt{(\gamma/2)^2-U^2+iU\gamma\overline{\xi}}\ .
\end{equation}
The result (\ref{fX}) is sufficient for calculating the average transmission, Eq. (\ref{TTT}), and the average dot occupation, Eq. (\ref{QQd}).
Both quantities are sums of two Lorentzians, with peaks related to the two poles $\Gamma+i\omega=s^{}_\pm$. Specifically, at $\overline{\xi}=0$ $u$ is real (or imaginary) for $\gamma>2U$ (or $\gamma<2U$), and the two peaks are centered at the same energy $\omega=U\overline{\xi}$, with widths $(\Gamma+\gamma/2\mp u)$ (or at $\omega=U\overline{\xi}\mp |u|$, with the same width $\Gamma+\gamma/2$) \cite{we2016}. When $\overline{\xi}\ne 0$ there are always two separate peaks, but they also overlap at small $U$.

It turns out that all the energy currents require the average of the product
\begin{equation}
Y(t,t'',t')=\epsilon^{}_d(t)X(t'',t')=U[\xi(t)-\overline{\xi}]X(t'',t')\ ,
\end{equation}
for $t\geq t''>t'$, where we remembered that we set the zero of energy at $\epsilon^{}_0=\epsilon+U\overline{\xi}$.
In matrix form, the conditional averages of $Y$ are found from
\begin{equation}
{\bf Y}(t,t'',t')=U{\bf X}(t'',t'){\bf P}(t-t'')(\sig^{}_z-\overline{\xi}{\bf I})\ .
\label{YYY}
\end{equation}

Specifically, the  energy  currents
 require the average of $\epsilon^{}_d(t)I^{}_L(t)$
and of $\epsilon^{}_d(t)I^{}_d(t)$, which contain
 $Y^{}_1(t,t')=Y(t,t,t')$.
In matrix form,
\begin{equation}
{\bf Y}^{}_1(t,t')={\bf Y}(t,t,t')=U{\bf X}(t-t')(\sig^{}_z-\overline{\xi}{\bf I})\ ,
\end{equation}
with the Laplace transform
\begin{equation}
\widetilde{\bf Y}^{}_1(s)=U\widetilde{\bf X}(s)(\sig^{}_z-\overline{\xi}{\bf I})\ .
\end{equation}
Averaging and some algebra yield
\begin{eqnarray}
\overline{\widetilde{ Y}}^{}_1(s)={\rm Tr}[{\bf T}\widetilde{\bf Y}^{}_1(s)]
=-i\overline{{\cal E}^2}{\cal F}^a(\omega)\equiv -\frac{i\overline{{\cal E}^2}}{[\omega-i\Gamma][\omega+2U\overline{\xi}-i(\Gamma+\gamma)]-\overline{{\cal E}^{2}}}\ ,
\end{eqnarray}
with the ``effective electric field"  (squared),
\begin{equation}
\overline{{\cal E}^{2}}= U^{2}(1-\overline{\xi}^{2})=4U^{2}p^{}_{+}(T)p^{}_{-}(T)\ .
\label{E2}
\end{equation}
As we show below, this quantity, which vanishes when $T=0$ [see Eq. (\ref{overx})] also determines the power absorbed by the junction.

 The calculation of $\epsilon^{}_d(t)G^<_{dd}(t,t)$, which is required for calculating $\overline{I^E_L}$ and $\overline{I^E_{{\rm tun},L}}$, see Eqs. (\ref{GDless}), (\ref{ILEWt}) and (\ref{IRtun4}), requires the average of $\epsilon^{}_d(t)K(t,\omega)$.
The conditional averages of this product become
\begin{eqnarray}
& \int_{-\infty}^t dt^{}_1 e^{-2\Gamma(t-t^{}_1)}\int_0^\infty d\tau e^{-(\Gamma+i\omega) \tau}{\bf Y}(t,t^{}_1,t^{}_1-\tau)\nonumber\\
&=U\int_{-\infty}^t dt^{}_1 e^{-2\Gamma(t-t^{}_1)}\int_0^\infty d\tau e^{-(\Gamma+i\omega)\tau}{\bf X}(t^{}_1,t^{}_1-\tau){\bf P}(t-t^{}_1)[\sig^{}_z-\overline{\xi}{\bf I}]\ .
\label{epsGless1}
\end{eqnarray}
 The integration over $\tau$ yields $\widetilde{\bf X}(\Gamma+i\omega)$, independent of  $t^{}_1$,
 and then the integration over $t^{}_1$ yields $\widetilde{\bf P}(2\Gamma)$. The resulting matrix is therefore
 $U\widetilde{\bf X}(\Gamma+i\omega)\widetilde{\bf  P}(2\Gamma)[\sig^{}_z-\overline{\xi}{\bf I}]$, 
with the average
\begin{equation}
- \frac{i\overline{{\cal E}^2}}{2\Gamma+\gamma}{\cal F}^a(\omega)\ .
 \end{equation}
 Thus,
 \begin{equation}
\overline{ i\epsilon_d(t)G^<_{dd}(t,t)}=\frac{\overline{{\cal E}^2}}{2\Gamma+\gamma}\int\frac{d\omega}{2\pi}\Sigma^<(\omega)[{\cal F}^a(\omega)-cc]\ .
\end{equation}

Finally, the calculation of
$P^{}_d$, Eq. (\ref{P}), requires the average of
\begin{equation}
\big[\frac{d}{dt}\epsilon^{}_d(t)\big] X(t^{}_1,t^{}_1-\tau)\equiv \frac{\partial}{\partial t}Y(t,t^{}_1,t^{}_1-\tau)\ .
\end{equation}
Using
 Eq. (\ref{dP}), the matrix form for this becomes
\begin{equation}
U{\bf X}(t^{}_1,t^{}_1-\tau){\bf W}{\bf P}(t-t^{}_1)[\sig^{}_z-\overline{\xi}{\bf I}]\ .
\end{equation}
To calculate the conditional averages of $P^{}_d$ we need the integral
\begin{eqnarray}
&\int_{-\infty}^t dt^{}_1 e^{-2\Gamma(t-t^{}_1)}\int_0^\infty d\tau e^{-(\Gamma+i\omega) \tau} \frac{\partial}{\partial t}{\bf Y}(t,t^{}_1,t^{}_1-\tau)\nonumber\\
&=U\int_{-\infty}^t dt^{}_1 e^{-2\Gamma(t-t^{}_1)}\int_0^\infty d\tau e^{-(\Gamma+i\omega)\tau}{\bf X}(t^{}_1,t^{}_1-\tau){\bf W}{\bf P}(t-t^{}_1)[\sig^{}_z-\overline{\xi}{\bf I}]\nonumber\\
&=U\widetilde{X}(\Gamma+i\omega){\bf W}\widetilde{\bf P}(2\Gamma)[\sig^{}_z-\overline{\xi}{\bf I}]\ .
\label{epsGless2}
\end{eqnarray}
Multiplication of the matrices and averaging finally gives
\begin{equation}
\frac{i\gamma\overline{{\cal E}^2}}{2\Gamma+\gamma}{\cal F}^a(\omega)\ .
\end{equation}
Therefore,
\begin{equation}
\overline{P^{}_d}=-i\overline{G^{<}_d(t,t)[\partial\epsilon^{}_d/\partial t]}=
-\frac{4\gamma\overline{{\cal E}^2}}{2\Gamma+\gamma}\int\frac{d\omega}{2\pi}[\Gamma^{}_Lf^{}_L(\omega)+\Gamma^{}_Rf^{}_R(\omega)]{\rm Im}[{\cal F}^a(\omega)]\ .
\end{equation}
Combining all these results into Eq. (\ref{ILEWt}) yields Eq. (\ref{IELW}). Inserting these results into Eq. (\ref{IRtun4}) yields $\overline{I^E_{{\rm tun},L}}=0$.

\section{Periodic electric field}\label{EE}

In this case, the dot energy is given by Eq. (\ref{edosc}).
Setting $\epsilon$ as the zero of energies, $X(t,t')$ becomes
\begin{equation}
X(t,t')=e^{iU\int_{t'}^t d\tau\cos(\Omega\tau)}=e^{ix[\sin(\Omega t)-\sin(\Omega t')]}=\sum_{n,m=-\infty}^\infty J^{}_n(x)J^{}_m(x)e^{i\Omega(n t-mt')}\ ,
\label{Xper}
\end{equation}
where $J^{}_n(x)$ is the Bessel function of the first kind and $x\equiv U/\Omega$.

In the present case, averaging is done over one period of the oscillating field,
\begin{equation}
\overline{Z(t,t')}=\frac{\Omega}{2\pi}\int_0^{2\pi/\Omega} dt Z(t,t')\ ,
\end{equation}
and therefore
\begin{equation}
\overline{X(t,t-\tau)}=\sum_{n=-\infty}^\infty [J^{}_n(x)]^2e^{in\Omega \tau}\ .
\end{equation}
Indeed, this average depends only on the time difference $\tau=t-t'$.
The Laplace transform (\ref{LT}) of this average is now
\begin{equation}
\widetilde{\overline{X}}(s)=\sum_{n=-\infty}^\infty \frac{[J^{}_n(x)]^2}{s-in\Omega}\ .
\end{equation}
Therefore, Eq. (\ref{LTGa}) yields
\begin{equation}
\overline{G^a_{dd}(\omega)}=i\sum_{n=-\infty}^\infty \frac{[J^{}_n(x)]^2}{\Gamma+i(\omega-n\Omega)}\ ,
\end{equation}
and [Eq. (\ref{TTT}), (\ref{QQd})]
\begin{equation}
{\cal T}(\omega)=\frac{4\Gamma^{}_L\Gamma^{}_R}{\Gamma}{\rm Im}\overline{[G^a_{dd}(\omega)]}=\sum_{n=-\infty}^\infty [J^{}_n(x)]^2 \frac{4\Gamma^{}_L\Gamma^{}_R}{\Gamma^2+(\omega-n\Omega)^2}\ ,
\end{equation}
\begin{equation}
\overline{Q^{}_d}=2\int\frac{d\omega}{2\pi}[\Gamma^{}_Lf^{}_L(\omega)+\Gamma^{}_Rf^{}_R(\omega)]\sum_{n=-\infty}^\infty \frac{[J_n(x)]^2}{\Gamma^2+(\omega-n\Omega)^2}\ .
\label{QQd2}
\end{equation}
The average transmission is thus a sum over many Breit-Wigner peaks, at energies $\omega=n \Omega$, all with the same width $\Gamma$. The charge on the dot grows whenever $\omega$ crosses one of thee peaks, ading another contribution to the integral in Eq. (\ref{QQd2}).

The remaining physical quantities require the time-dependent $Q^{}_d(t)$. From Eq. (\ref{GDless}) we have
\begin{equation}
Q^{}_d(t)=-iG^<_{dd}(t,t)=2\int\frac{d\omega}{2\pi}[\Gamma^{}_Lf^{}_L(\omega)+\Gamma^{}_Rf^{}_R(\omega)][K(t,\omega)+cc]\ ,
\end{equation}
with $K(t,\omega)$ from Eq. (\ref{KKK}).
Substituting Eq. (\ref{Xper}), we find
\begin{equation}
K(t,\omega)=\sum_{n,m=-\infty}^\infty J^{}_n(x)J^{}_m(x)\frac{e^{i(n-m)\Omega t}}{[\Gamma+i(\omega-m\Omega)][2\Gamma+i(n-m)\Omega]}\ .
\label{Kt}
\end{equation}

The average power, $\overline{P_d(t)}$, is thus
\begin{equation}
\overline{P^{}_d}=2\int\frac{d\omega}{2\pi}[\Gamma^{}_Lf^{}_L(\omega)+\Gamma^{}_Rf^{}_R(\omega)]\Big[\overline{\frac{d\epsilon^{}_d(t)}{dt}K(t,\omega)}+cc\Big]\ ,
\end{equation}
with
\begin{eqnarray}
&\overline{\frac{d\epsilon^{}_d(t)}{dt}K(t,\omega)}=\frac{i\Omega U}{2}\overline{(e^{i\Omega t}-e^{-i\Omega t})K(t,\omega)}=\nonumber\\
&-\frac{i\Omega U}{2}\sum_{m=-\infty}^\infty \frac{J^{}_m(x)}{\Gamma+i(\omega-m\Omega)}\Big[\frac{J^{}_{m+1}(x)}{2\Gamma+i\Omega}-\frac{J^{}_{m-1}(x)}{2\Gamma-i\Omega}\Big]\ ,
\end{eqnarray}
hence
\begin{eqnarray}
&\overline{P^{}_d}=-\frac{4U\Omega\Gamma}{4\Gamma^2+\Omega^2}\int\frac{d\omega}{2\pi}[\Gamma^{}_Lf^{}_L(\omega)+\Gamma^{}_Rf^{}_R(\omega)]\nonumber\\
&\times \sum_{m=-\infty}^\infty \frac{J^{}_m(x)}{\Gamma^2+(\omega-m\Omega)^2}\big[m\Omega J^{}_m(x)/x-2(\omega-m\Omega)J'_m(x)\big]\ ,
\label{Pdavper}
\end{eqnarray}
where we used $J^{}_{m+1}(x)+J^{}_{m-1}(x)=2mJ^{}_m(x)/x$, $J^{}_{m+1}(x)-J^{}_{m-1}(x)=-2J'_m(x)$.

For simplicity, we present results for $f_L(\omega)=f_R(\omega)=f(\omega)$.
Integration by parts over $\omega$ we have
\begin{equation}
\int_{-\infty}^\infty d\omega f(\omega)F'(\omega)=-F(-\infty)+\int_{-\infty}^\infty d\omega [-f'(\omega)]F(\omega)\ ,
\end{equation}
where we need $F^{}_1(\omega)=\arctan[(\omega-m \Omega)/\Gamma]/\Gamma$ and $F^{}_2=\ln[\Gamma^2+(\omega-m \Omega)^2]/2$.
Although $F^{}_2(\omega)$ diverges in the limit $\omega\rightarrow -\infty$, this term multiplies the sum $\sum_m J_m(x)J'_m(x)=0$, since $\sum_m J_m(x)J_{m+1}(x)=\sum_m J_m(x)J_{m-1}(x)$. Similarly, $\sum_m mJ_m(x)^2=0$. Therefore, we ignore the contributions from $\omega\rightarrow -\infty)$.
At low reservoirs temperatures, $[-f'(\omega)]$ is close to $\delta(\omega)$. Therefore, we approximate the results by setting $T=0$, which allows analytic expressions.
Using also the common chemical potential $\mu^{}_L=\mu^{}_R=\mu$, the average incoming energy flux becomes
\begin{eqnarray}
&\overline{P^{}_d}=-\frac{4U\Omega\Gamma^2}{4\Gamma^2+\Omega^2}
 \sum_{m=-\infty}^\infty J^{}_m(x)\Big(\frac{m\Omega }{\Gamma x}J^{}_m(x)\arctan[(\mu-m \Omega)/\Gamma]\nonumber\\
&-J'_m(x)\ln[\Gamma^2+(\mu-m \Omega)^2]\Big)/(2\pi)\ .
\label{Pdfin}
\end{eqnarray}

Similarly, the average of $\epsilon^{}_d(t)Q^{}_d(t)$ requires
\begin{eqnarray}
&\overline{\epsilon^{}_d(t)K(t,\omega)}=\frac{ U}{2}\overline{(e^{i\Omega t}+e^{-i\Omega t})K(t,\omega)}=\nonumber\\
&\frac{ U}{2}\sum_{m=-\infty}^\infty \frac{J^{}_m(x)}{\Gamma+i(\omega-m\Omega)}\Big[\frac{J^{}_{m+1}(x)}{2\Gamma+i\Omega}+\frac{J^{}_{m-1}(x)}{2\Gamma-i\Omega}\Big]\ ,
\end{eqnarray}
and therefore
\begin{eqnarray}
&\overline{\epsilon^{}_d(t)[K(t,\omega)+cc]}=\nonumber\\
&\frac{2U}{4\Gamma^2+\Omega^2}\sum_{m=-\infty}^\infty \frac{J^{}_m(x)}{\Gamma^2+(\omega-m\Omega)^2}\big[2m\Gamma^2 J^{}_m(x)/x +\Omega(\omega-m\Omega)J'_m(x)\big]\ .
\end{eqnarray}
Combining this with Eqs. (\ref{ILEWt}), (\ref{Pdavper}) and (\ref{TTT})
finally yields Eq. (\ref{IELW}) and  $I^E_{{\rm tun},L}=0$.

\section{Results and Discussion}\label{dis}

\begin{figure}[htp]
\includegraphics[width=5.4cm]{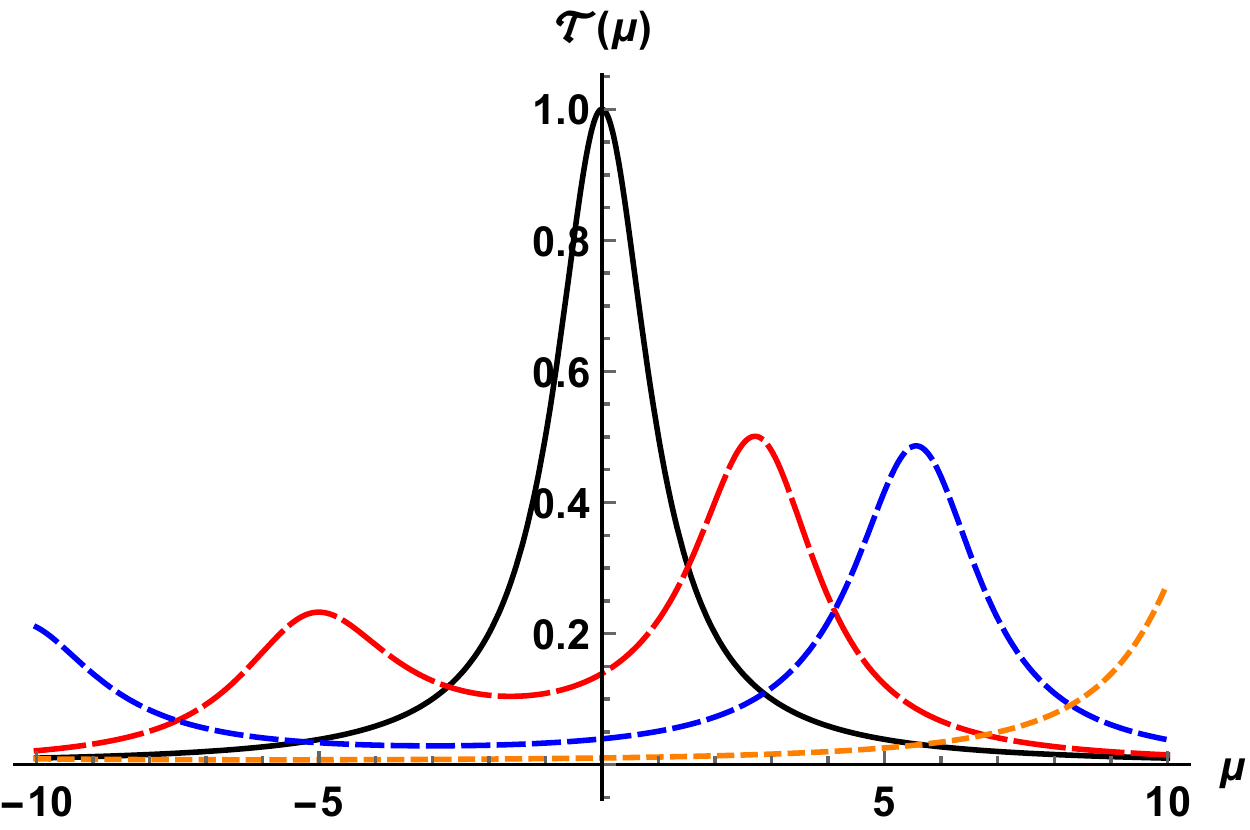}\ \ \ \ \ \ \ \ \ \includegraphics[width=5.4cm]{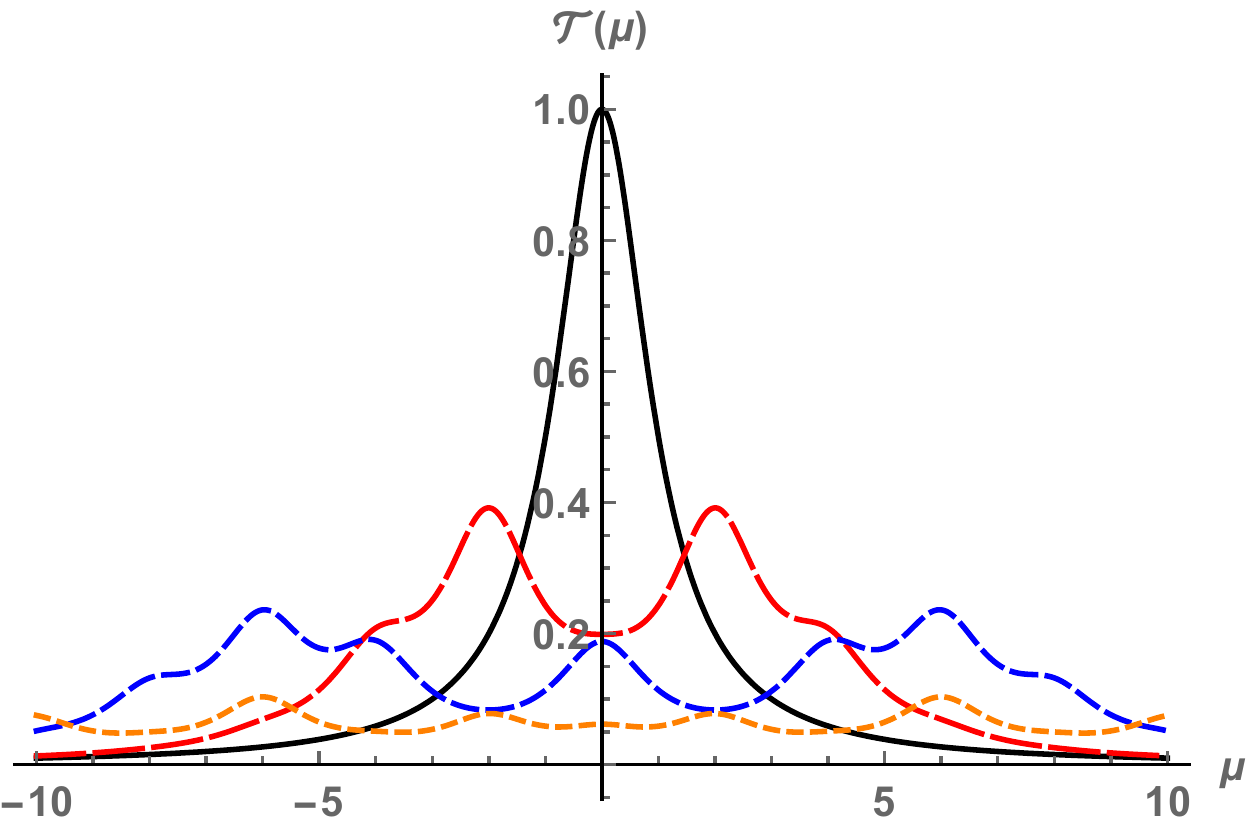}
\caption{The average transmission ${\cal T}$ for identical electronic reservoirs (see text), versus the common chemical potential $\mu$. Energies are in units of $\Gamma$. Left: TN model, with  $\overline{\xi}=0.3$ (that is a finite effective fluctuator temperature $T$) and jump rate  $\gamma=1$.  Right: Oscillating field, with oscillaion frequency $\Omega=2$.  The different curves with decreasing dashes are for $U=0,~,4,~8,~16$ ($U$ is the amplitude of the fluctuating or oscillating energy level on the dot).  }
\label{Tmu}
\end{figure}

\begin{figure}[htp]
\includegraphics[width=5.4cm]{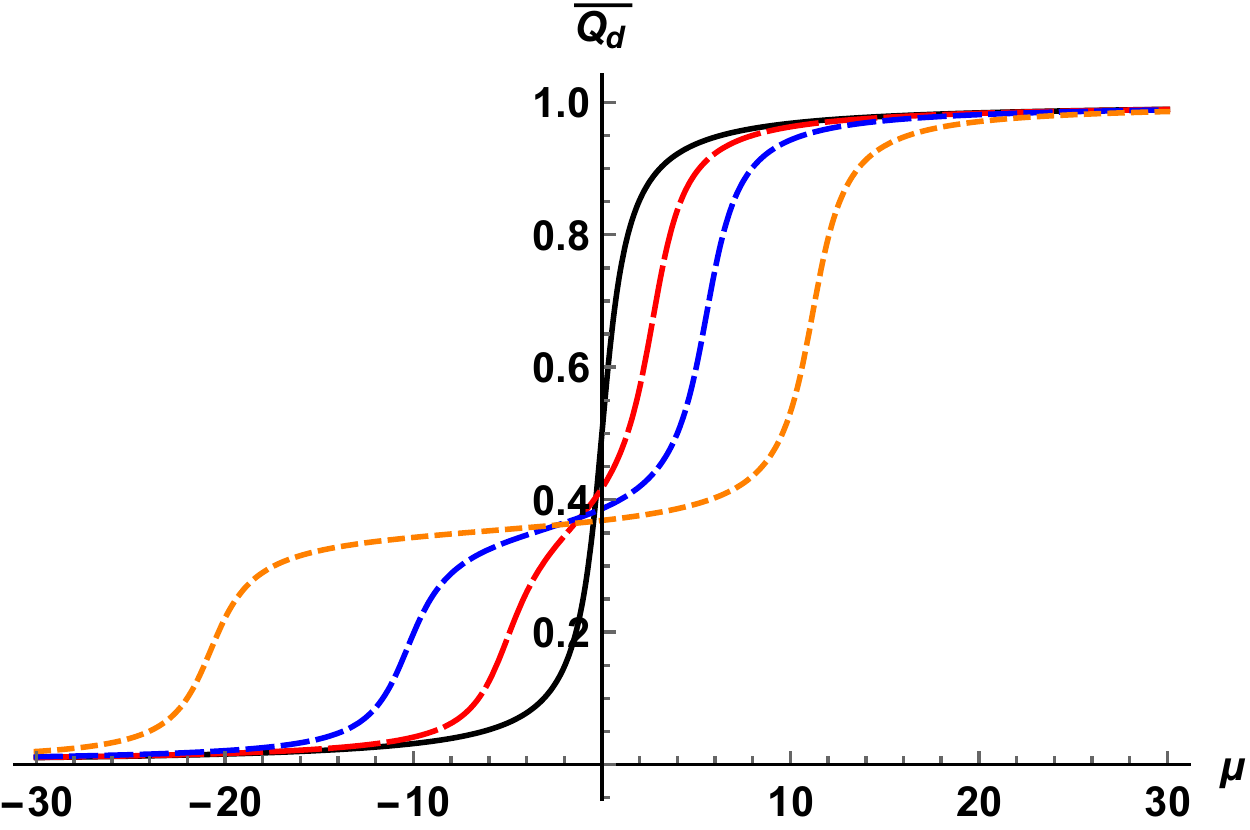}\ \ \ \ \ \ \ \ \ \includegraphics[width=5.4cm]{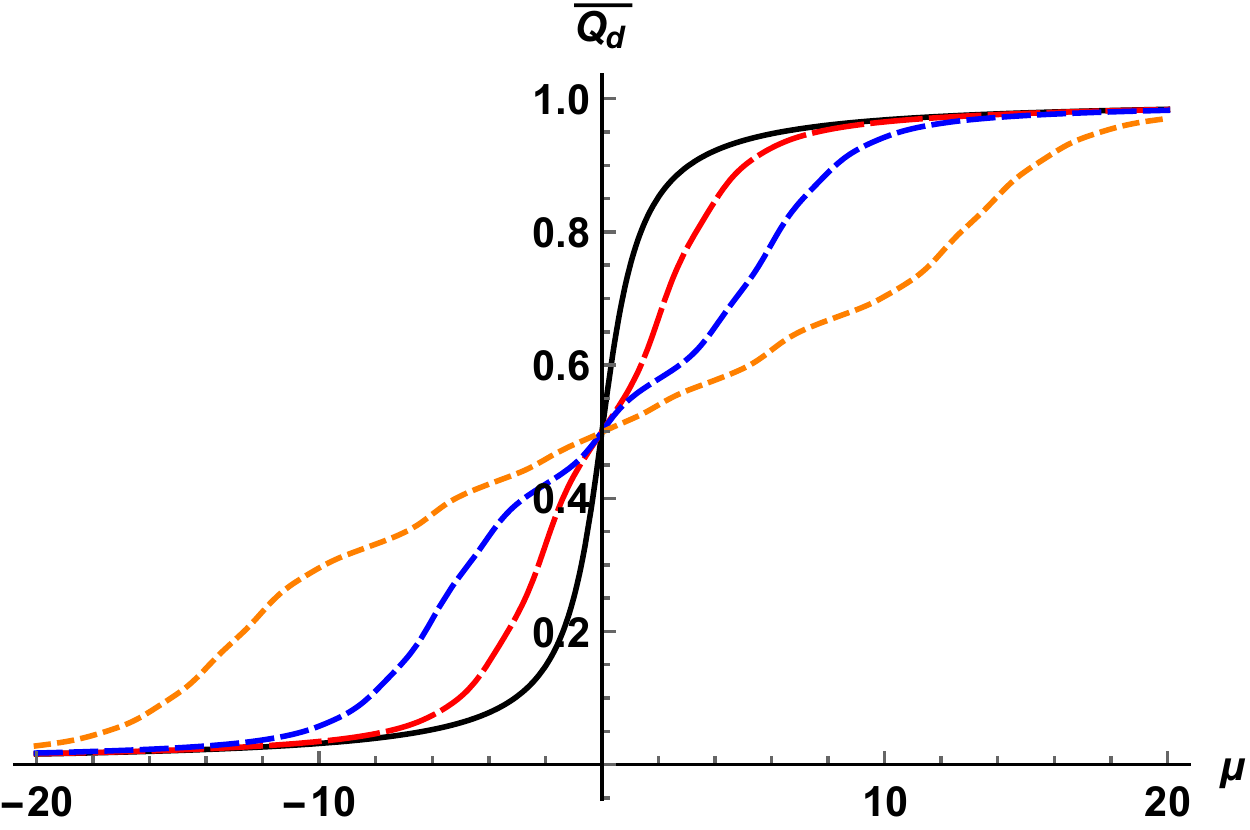}
\caption{The average charge on the dot, $\overline{Q^{}_d}$, versus $\mu$, for the same models and parameters as in Fig. \ref{Tmu}.  }
\label{Qdmu}
\end{figure}

\begin{figure}[htp]
\includegraphics[width=5.4cm]{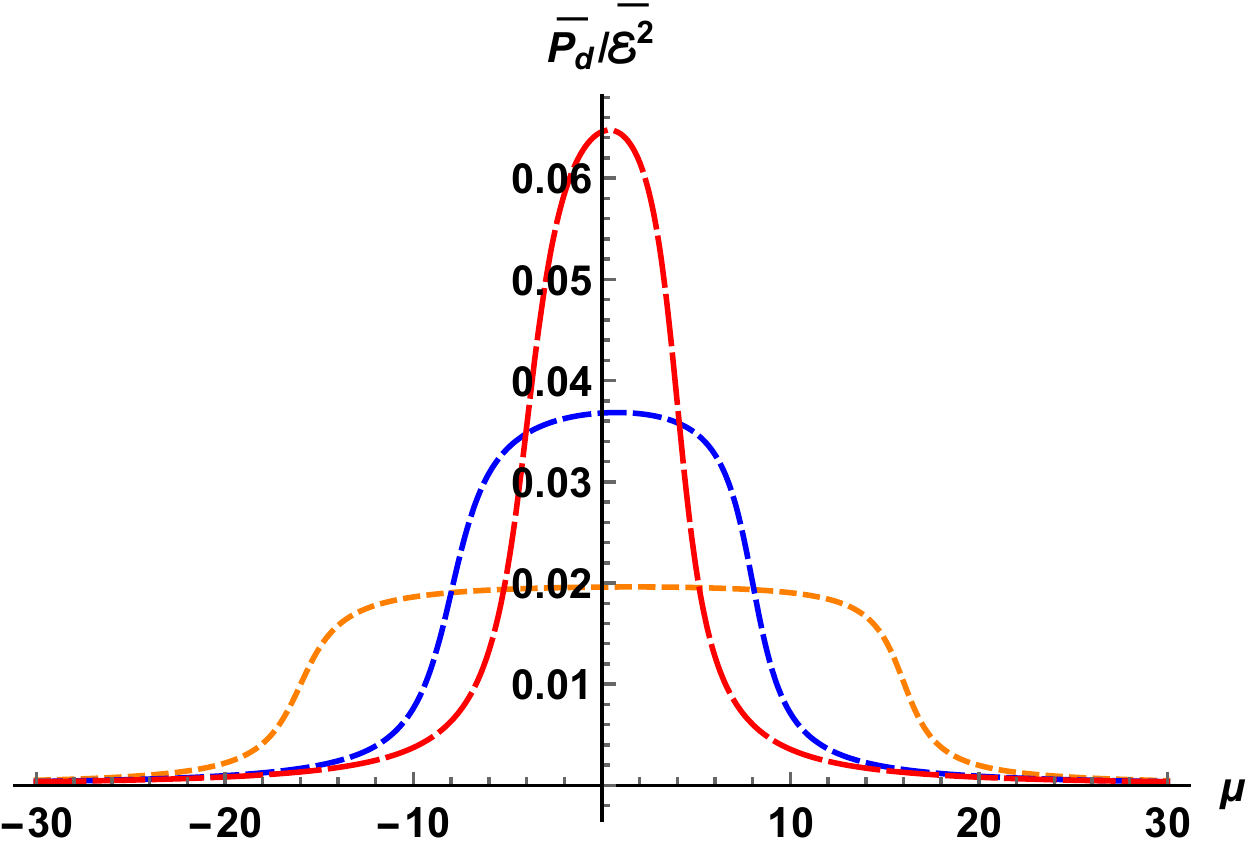}\ \ \ \ \ \ \ \ \ \includegraphics[width=5.4cm]{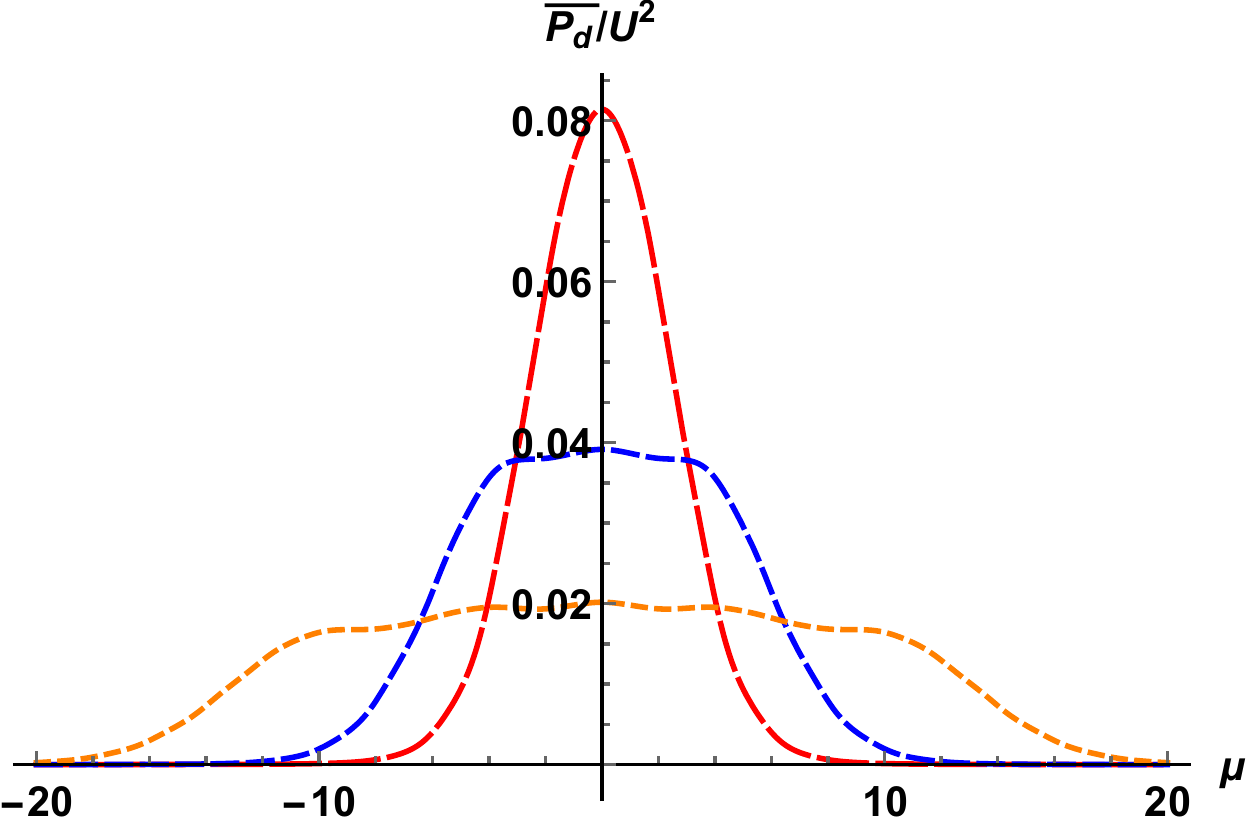}
\caption{The average power absorbed by the junction $\overline{P^{}_d}$ (in units of the effective electric field (squared) $\overline{{\cal E}^{2}}$, for the same models and parameters as in Fig. \ref{Tmu}.  }
\label{Pdmu}
\end{figure}

Figures \ref{Tmu}, \ref{Qdmu} and \ref{Pdmu} compare results for the average transmission ${\cal T}(\mu)$, average charge on the dot $\overline{Q_d}(\mu)$ and average energy flux from the time-dependent electric field $\overline{P_d}(\mu)$ for the telegraph noise (TN) model (left) and the oscillating field model (right). All the energies are in units of the bare (without the effective time-dependent field) resonance width $\Gamma$. In all the figures we used a symmetric dot, $\Gamma^{}_L=\Gamma^{}_R=\Gamma/2$, and calculated the results in the limit of zero reservoirs temperatures. As explained, the deviations of $-\partial f/\partial\omega$ from $\delta(\omega)$ at low temperatures are very small.

In the TN model, for $U\geq\gamma/2$ the transmission shows two peaks. Since the plot was calculated with $\overline{\xi}=.3$, the two peaks are not symmetric. As the effective fluctuator temperature $T$ is decreased, $\overline{\xi}$ increases, and the right (left) hand side peak grows (decreases). The corresponding charge on the dot shows two steps, as the chemical potential crosses the above two peaks.  The average incoming energy flux, $\overline{P^{}_d}$, shows a broad (slightly asymmetric) peak, with a maximum near the average dot energy.
The oscillating field model yields similar results, but with the peaks and the steps occurring at all the harmonics of $\Omega$.
Basically, the two models are qualitatively similar. As stated in the Introduction, this similarity raises questions on the utility of the TN to represent the random environment of the system.

Apparently, the TN model yields an unphysical result: the fluctuator generates an energy current into the electronic reservoirs, irrespective of its effective temperature. As mentioned, this result is probably due to the lack of back action from the electronic system to the fluctuator's heat bath.
The full coupling of a molecular junction to a phonon bath has been the topic of extensive research \cite{3term}. In a simple model, our Eq. (\ref{Hsys}) is replaced by
\begin{equation}
{\cal H}^{}_{\rm sys+env} = [\epsilon^{}_0+U(b^\dagger_{}+b^{}_{})]d^\dagger_{}d^{}_{}+\omega^{}_0(b^\dagger_{}b^{}_{}+1/2)\ ,
\label{3T}
\end{equation}
where  $b^\dagger_{}$ ($b^{}_{}$) creates (annihilates) an excitation of the (single) harmonic oscillator, of frequency  $\omega^{}_0$.
The electron-phonon coupling  allows inelastic processes, in which the electrons exchange energy with the vibrational mode, and these generate energy currents between the phononic heat bath and the electrons.
In our calculations, the electrons never undergo such inelastic processes, and therefore these calculations do not imitate the quantum models of molecular junctions.

A generalization of Eq. (\ref{3T}) was discussed in Ref. \cite{pramana}:
\begin{equation}
{\cal H}^{}_{\rm sys+env}=U d^\dagger_{}d \tau^{}_z + (E^{}_0/2) \tau^{}_z  + \sum_q G^{}_{\bf q} \tau^{}_x (b^{}_{\bf q} + b^\dagger_{\bf q}) + \sum_q \omega^{}_{\bf q} b^\dagger_{\bf q} b^{}_{\bf q} \ .
 \label{SD}
\end{equation}
The last three terms in Eq. (\ref{SD}) describe phonon-assisted tunnelling of a defect in an asymmetric (parameterized by the asymmetry energy $E^{}_0$). The interaction term $G^{}_{\bf q}$ facilitates tunnelling (via the pseudo Pauli spin $\tau^{}_x$ which is completely off-diagonal in the representation in which $\tau^{}_z$ is diagonal). Evidently the eigenstates of $\tau^{}_z$ mimic the two minima of the well. The tunnelling is accentuated by phonon annihilation and creation operators $b^{}_{\bf q}$ and $b^\dagger_{\bf q}$, which are further tickled by the free phonon term with energy $\omega^{}_{\bf q}$. The physics behind Eq. (\ref{SD}) is clear: $\tau^{}_x$ flips $\tau^{}_z$ between $+1$ and $-1$, thus changing the sign of the $U$ term and simultaneously moving (tunnelling) the defect from one minimum of the well to another. This tunnelling can however be incoherent because the phonon operators are time-fluctuating in the interaction picture of the free phonon term.
It is this incoherent tunnelling that is envisaged in our model to be a TN, as detailed in the text. Thus the TN is physically motivated, yet simple enough to carry out exact calculations for what a mesoscopic environment does. There is one caveat though. The TN model, while appearing to be reasonable for computing the charge current between the dot and the leads, yields a seemingly unphysical result when it comes to the energy current. This is especially irksome when the temperature of the phonon bath $T$ is brought below $T_L$ and $T_R$, persisting down to $T = 0$.
This fallacy can be traced to an artefact of the TN model. As  shown in Ref. \cite{pramana}, the route to the TN from Eq. (\ref{SD}) is smoothened by several approximations – neglect of $U$ in the time-evolution, assumption of Markovianness by ignoring the frequency-dependence of the self-energy and above all, the stipulation that the phonon bath can be treated as classical, at a temperature higher than most other energy scales in the problem. These assumptions are well-documented in the literature on defect kinetics wherein tunnelling at low temperatures goes over to activated diffusion across an Arrhenius barrier, at high temperatures \cite{pramana}. Thus it seems that the applicability of the TN model, as far as a mesoscopic environment is concerned, is dubious when $T$ falls below both $T_L$ and $T_R$. For a complete treatment one may carry out a fully self consistent calculation by juxtaposing the Hamiltonian in Eq. (\ref{SD}) with the quantum Hamiltonian of the dot, coupled with the Fermionic leads via the tunnelling terms (See Eqs. (\ref{Htun}) and (\ref{Hleads}) of the text). Such a full-fledged treatment is relegated to the future.

\begin{acknowledgements}

We thank  Shmuel Gurvitz  for  discussions.
This work was supported by the Israeli Science Foundation (ISF),  by the infrastructure program of Israel
Ministry of Science and Technology under contract
3-11173, and by a grant from the Pazy foundation. SD is grateful to the Indian National Science Academy for support through its Senior Scientist scheme.
\end{acknowledgements}

\appendix

\section{Dyson equations for the Green's functions}

\setcounter{equation}{0}%
\renewcommand\theequation{A.\arabic{equation}}%

The Green's functions in Eq. (\ref{IL}) are derived from the Dyson equations,
  \begin{equation}
 G^{}_{{\bf k}d}(t,t')=\int dt^{}_{1}g^{}_{\bf k}(t,t^{}_{1})V^{}_{L}G^{}_{dd}(t^{}_{1},t')\ ,\ \ G^{}_{d{\bf k}}(t,t')=\int dt^{}_{1}G^{}_{dd}(t,t^{}_{1})V^{\ast}_{L}g^{}_{\bf k}(t^{}_{1},t')
\ ,
\label{DysonGkd}
\end{equation}
where $g^{}_{\bf k}(t,t^{}_1)$ is the decoupled Green's function in the lead.  Here and below, all the unspecified time integrals begin at $t^{}_0=-\infty$ and end at $t$. Substituting Eq. (\ref{DysonGkd}) into Eq. (\ref{IL}),  we have
\begin{equation}
I^{}_{L}(t)
=\int dt^{}_{1}[\Sigma^{}_{L}(t,t^{}_{1})G^{}_{dd}(t^{}_{1},t)-G^{}_{dd}(t,t^{}_{1})\Sigma^{}_{L}(t^{}_{1},t)]^{<}_{}\ ,
\label{ILd}
\end{equation}
where $\Sigma^{}_{L}(t,t')$ is the self energy due to the tunnel coupling with the left lead,
\begin{equation}
\Sigma^{}_{L}(t,t')=|V^{}_{L}|^{2}g^{}_{L}(t,t')\equiv |V^{}_{L}|^{2}\sum_{\bf k} g^{}_{\bf k}(t,t^{}_1) \ .
\label{SigL}
\end{equation}
Noting that
\begin{equation}
g^{r(a)}_{\bf k}(t-t')=\mp i\Theta (\pm t\mp t')\langle\{ c^{}_{\bf k}(t),c^{\dagger}_{\bf k}(t')\}\rangle
=\mp i\Theta (\pm t \mp t')e^{-i\epsilon^{}_{k}(t-t')}\ ,
\label{gr}
\end{equation}
and
\begin{equation}
g^{<}_{\bf k}(t-t')=if(\epsilon^{}_{k})e^{-i\epsilon^{}_{k}(t-t')}\ ,\ \ \ f^{}_L(\epsilon^{}_{k})=\langle c^{\dagger}_{\bf k}c^{}_{\bf k}\rangle
\ ,
\label{gl}
\end{equation}
[The superscript $r(a)$  indicates
the  retarded (advanced) Green's  function and corresponds to the upper (lower) sign on the right hand-side, while
\begin{equation}
f^{}_{L(R)}(\omega)=[e^{(\omega-\mu^{}_{L(R)})/(k^{}_{\rm B}T^{}_{L(R)})}+1]^{-1}
\label{Fermi}
\end{equation}
is the Fermi distribution in the left (right) lead],
and using the wide-band limit, in which the densities of states in the reservoirs are assumed to be independent of the energy \cite{Jauho1994}, we have
 \begin{eqnarray}
 \Sigma^{r(a)}_{L}(t,t')=\mp i\Gamma^{}_{L}\delta(t-t')\ ,\ \ \ \
 \Sigma^<_{L}(t,t')=2i\Gamma^{}_{L}\int\frac{d\omega}{2\pi}e^{-i\omega(t-t')}f^{}_{L}(\omega)\ .
\label{SIG}
 \end{eqnarray}

Applying the Langreth rule, $[AB]^<=A^rB^<+A^<B^a$ \cite{Jauho1994}, to Eq. (\ref{ILd}), and substituting Eqs. (\ref{SIG}), we find
\begin{equation}
I^{}_L(t)=2i\Gamma^{}_L\Big(\int\frac{d\omega}{2\pi}f^{}_L(\omega)\big[e^{-i\omega(t-t^{}_1)}G^a_{dd}(t^{}_1,t)-e^{-i\omega(t^{}_1-t)}G^r_{dd}(t,t^{}_1)\big]
-G^<_{dd}(t,t)\Big)\ .
\label{IL0}
\end{equation}
The time-dependent particle flux into the right lead is derived from these equations by interchanging $L\Leftrightarrow R$ and ${\bf k}\Leftrightarrow {\bf p}$.

The Dyson equation for the  Green's functions on the dot reads
\begin{equation}
G^{}_{dd}(t,t')
=g^{}_{d}(t,t')+\int \int dt^{}_{1}dt^{}_{2}g^{}_{d}(t,t^{}_{1})\big[V^\ast_L G^{}_{Ld}(t^{}_1,t')+V^\ast_R G^{}_{Rd}(t^{}_1,t')\big]\ ,
\label{Gdd1}
\end{equation}
where
 \begin{equation}
 G^{}_{Ld}(t,t')=\sum_{\bf k}G^{}_{{\bf k}d}(t,t')\ ,\ \ \  G^{}_{Rd}(t,t')=\sum_{\bf p}G^{}_{{\bf p}d}(t,t')\ ,
 \end{equation}
while  $g^{}_d$ is the Green's function on the isolated dot, with
\begin{equation}
g^{r(a)}_{d}(t,t')=\mp i\Theta (\pm t\mp t')e^{ -i\int_{t'}^{t}dt^{}_{1}\epsilon^{}_{d}(t^{}_{1})}\ ,
\label{grad}
\end{equation}
and $g^{<}_{d}=0$, since it is assumed that the dot is empty in the decoupled junction.
Inserting Eq. (\ref{DysonGkd}) into Eq. (\ref{Gdd1}) yields
\begin{equation}
G^{}_{dd}(t,t')
=g^{}_{d}(t,t')+\int \int dt^{}_{1}dt^{}_{2}g^{}_{d}(t,t^{}_{1})\Sigma^{}_{}(t^{}_{1},t^{}_{2})G^{}_{dd}(t^{}_{2},t')\ ,
\label{DG}
\end{equation}
where
\begin{equation}
\Sigma(t,t')=\Sigma^{}_{L}(t,t')+\Sigma^{}_{R}(t,t')\ .
\end{equation}

Using Eqs. (\ref{SIG}), Eq. (\ref{Gdd1}) becomes
\begin{equation}
G^{r(a)}_{dd}(t,t')=g^{r(a)}_{d}(t,t')\mp i \Gamma\int dt^{}_1g^{r(a)}_d(t,t^{}_1)G^{r(a)}_{dd}(t^{}_1,t'),\ \ \ \ \Gamma=\Gamma^{}_L+\Gamma^{}_R\ ,
\end{equation}
with the solution given in Eq. (\ref{GDRA}).

Using the Langreth rules, the
 lesser Green's function on the dot obeys
\begin{equation}
G^<_{dd}(t,t')=\int dt^{}_1\int dt^{}_2 g^r_d(t,t^{}_1)\big[\Sigma^r(t^{}_1,t^{}_2)G^<_{dd}(t^{}_2,t')+\Sigma^<(t^{}_1,t^{}_2)G^a_{dd}(t^{}_2,t')\big]\ .
\end{equation}
Inserting Eqs. (\ref{SIG}), this becomes
\begin{equation}
G^<_{dd}(t,t')=-i\Gamma \int dt^{}_1 g^r_d(t,t^{}_1)G^<_{dd}(t^{}_1,t')+\int dt^{}_1\int dt^{}_2 g^r_d(t,t^{}_1)\Sigma^<(t^{}_1,t^{}_2)G^a_{dd}(t^{}_2,t')\big]\ .
\end{equation}
Differentiating this equation with respect to $t$, one finds
\begin{equation}
\frac{\partial G^<_{dd}(t,t')}{\partial t}=-i[\epsilon^{}_d(t)-i\Gamma]G^<_{dd}(t,t')-i\int dt^{}_1\Sigma^<(t,t^{}_1)G^a_{dd}(t^{}_1,t')\ ,
\label{dGless}
\end{equation}
with the solution
\begin{equation}
G^<_{dd}(t,t')=\int dt^{}_1\int dt^{}_2 G^r_{dd}(t,t^{}_1)\Sigma^<(t^{}_1,t^{}_2)G^a_{dd}(t^{}_2,t')\ .
\label{Glessdd}
\end{equation}
With Eqs. (\ref{GDRA}), and setting $t'=t$, this becomes
\begin{equation}
G^{<}_{dd}(t,t)=\int\frac{d\omega}{2\pi}\Sigma^{<}_{}(\omega)\int^{t}dt^{}_{1}\int^{t}dt^{}_{2}e^{\Gamma(t^{}_{1}+t^{}_{2}-2t)}\
 e^{i\omega(t^{}_{2}-t^{}_{1})}X(t^{}_1,t^{}_2)\ ,
\label{GLTTA}
\end{equation}
where [see Eq. (\ref{SIG})]
\begin{equation}
\Sigma^<(\omega)=\Sigma^{<}_L(\omega)+\Sigma^<_R(\omega)=2i[\Gamma^{}_Lf^{}_L(\omega)+\Gamma^{}_Rf^{}_R(\omega)]\ .
\end{equation}
Changing the double integration,
\begin{equation}
\int^{t} dt^{}_{1}\int^{t}dt^{}_{2}F(t^{}_{1},t^{}_{2})=
 \int^{t} dt^{}_{1}\int^{t^{}_{1}}dt^{}_{2}[F(t^{}_{1},t^{}_{2} )+F(t^{}_{2},t^{}_{1})]\ ,
\end{equation}
and changing variables, $t^{}_2\rightarrow \tau=t^{}_1-t^{}_2$, yield Eq. (\ref{GDless}).

We now turn to the energy fluxes. The energy flux into the left lead is given in Eq. (\ref{ILE})
In terms of the Green's functions on the dot, this energy flux reads
\begin{equation}
I^{E}_{L}(t)
=\int dt^{}_{1}[\Sigma^{E}_{L}(t,t^{}_{1})G^{}_{dd}(t^{}_{1},t)-G^{}_{dd}(t,t^{}_{1}\Sigma^{E}_{L}(t^{}_{1},t)]^{<}_{}\ ,
\label{ILEd}
\end{equation}
where $\Sigma^{E}_{L}(t,t')$ is 
\begin{equation}
\Sigma^{E}_{L}(t,t')=\sum_{\bf k}\epsilon^{}_{k}|V^{}_{\bf k}|^{2}g^{}_{\bf k}(t,t')\ .
\label{SigEL}
\end{equation}
In the wide-band limit,
the Fourier transforms of $\Sigma^E_L$ are
\begin{equation}
\Sigma^{E,r(a)}_L(\omega)=\mp i\omega \Gamma^{}_{L(R)}\ ,\ \ \ \Sigma^{E,<}_{L(R)}(\omega)=2i\omega\Gamma^{}_{L(R)}f^{}_{L(R)}(\omega)\ .
\end{equation}
Using the Langreth rule in Eq. (\ref{ILEd}), we have
\begin{eqnarray}
&I^{E}_{L}(t)=\int dt_{1}\Big[\Sigma_{L}^{E,r}(t,t_{1})G^{<}_{dd}(t_{1},t)+\Sigma_{L}^{E,<}(t,t_{1})G^{a}_{dd}(t_{1},t)\nonumber\\
&-G^{r}_{dd}(t,t_{1})\Sigma_{L}^{E,<}(t_{1},t)-G_{dd}^{<}(t,t_{1})\Sigma_{L}^{E,a}(t_{1},t)\Big]\ .
\end{eqnarray}
The second and third term give
\begin{eqnarray}\label{5}
&\int dt_{1}\Big[\Sigma_{L}^{E,<}(t,t_{1})\overline{G^{a}_{dd}(t_{1},t)}-\overline{G^{r}_{dd}(t,t_{1})}\Sigma_{L}^{E,<}(t_{1},t)\Big]\nonumber\\
&=2i\Gamma_{L}\int\frac{d\omega}{2\pi}\omega f_{L}(\omega)\Big(\overline{G^{a}_{dd}(\omega)}-\overline{G^{r}_{dd}(\omega)}\Big)\ .
\end{eqnarray}
The first and last terms yield
\begin{eqnarray}\label{2}
&\int dt_{1}[\Sigma^{E,r}_{L}(t,t_{1})G^{<}_{dd}(t_{1},t)-G^{<}_{dd}(t,t_{1})\Sigma^{E,a}_{L}(t_{1},t)]\nonumber\\
&=\Gamma_{L}\int d\tau \Big(\frac{\partial\delta(\tau)}{\partial\tau}G_{dd}^{<}(t-\tau,t)-G_{dd}^{<}(t,t-\tau)\frac{\partial\delta(\tau)}{\partial\tau}\Big)\nonumber\\
&= -\Gamma_{L}\int d\tau \delta(\tau)\frac{\partial}{\partial\tau} \Big(G_{dd}^{<}(t-\tau,t)-G_{dd}^{<}(t,t-\tau)\Big)\nonumber\\
&=\Gamma^{}_{L}\int d\tau\delta (\tau)\int dt_{1}\int dt_{2}\frac{\partial}{\partial \tau}\Big[G^{r}_{dd}(t-\tau,t_{1})\Sigma^{<}(t_{1},t_{2})G_{dd}^{a}(t_{2},t)\nonumber\\
&-G^{r}_{dd}(t,t_{1})\Sigma^{<}(t_{1},t_{2})G_{dd}^{a}(t_{2},t-\tau)\Big]\nonumber\\
&=-i\Gamma_{L}\int dt_{1}\Big(\Sigma^{<}(t,t_{1})G_{dd}^{a}(t_{1},t)+G_{dd}^{r}(t,t_{1})\Sigma^{<}(t_{1},t)\Big)\nonumber\\
&-2i\Gamma_{L}\epsilon_{d}(t)G_{dd}^{<}(t,t)\ ,
\end{eqnarray}
where the last two steps used Eqs. (\ref{Glessdd}) and (\ref{dGless}).
Finally, we find Eq. (\ref{ILEWt}).

The
 energy fluxes which result from the temporal variation of the (left and right) tunneling Hamiltonians, Eq. (\ref{ITLE})
(with an analogous expression for $I^{E}_{{\rm tun},R}$) require the Green's functions $G_{{\bf kp}}(t,t')$. Using the Dyson equations
\begin{eqnarray}
&G_{{\bf kp}}(t,t')=V_{{\bf k}}V_{{\bf p}}^{*}\int dt_{1}\int dt_{2}g_{{\bf k}}(t,t_{1})G_{dd}(t_{1},t_{2})g_{{\bf p}}(t_{2},t')\nonumber\\
&G_{{\bf pk}}(t,t')=V_{{\bf p}}V_{{\bf k}}^{*}\int dt_{1}\int dt_{2}g_{{\bf p}}(t,t_{1})G_{dd}(t_{1},t_{2})g_{{\bf k}}(t_{2},t')\ ,
\end{eqnarray}
the Lagreth rules and Eq. (\ref{SIG}) we find
\begin{eqnarray}\label{11}
&I^{E}_{{\rm tun},L}(t)-\epsilon^{}_{d}(t)I^{}_{L}(t)+I^{E}_{L}(t)
\nonumber\\
&=\int dt_{1}\int dt_{2}\Big[\Sigma_{L}(t,t_{1})G_{dd}(t_{1},t_{2})\Sigma_{R}(t_{2},t)-\Sigma_{R}(t,t_{1})G_{dd}(t_{1},t_{2})\Sigma_{L}(t_{2},t)\Big]^{<}\nonumber\\
&=\int dt_{1}\int dt_{2}\Big[\Sigma_{L}^{r}(t,t_{1})\Big(G_{dd}^{r}(t_{1},t_{2})\Sigma_{R}^{<}(t_{2},t)+G_{dd}^{<}(t_{1},t_{2})\Sigma_{R}^{a}(t_{2},t)
+G_{dd}^{a}(t_{1},t_{2})\Sigma_{R}^{a}(t_{2},t)\Big)\nonumber\\&-\Sigma_{R}^{r}(t,t_{1})\Big(G_{dd}^{r}(t_{1},t_{2})\Sigma_{L}^{<}(t_{2},t)
+G_{dd}^{<}(t_{1},t_{2})\Sigma_{L}^{a}(t_{2},t)\Big)+G_{dd}^{a}(t_{1},t_{2})\Sigma_{L}^{a}(t_{2},t)\Big)\Big]\nonumber\\
&=2\Gamma_{L}\Gamma_{R}\int \frac{d\omega}{2\pi}\Big(f_{R}(\omega)-f_{L}(\omega)\Big)\int dt_{1}\Big[e^{-i\omega(t_{1}-t)}G_{dd}^{r}(t,t_{1})+e^{-i\omega(t-t_{1})}G_{dd}^{a}(t_{1},t)\Big]\ ,
\end{eqnarray}
Hence
\begin{eqnarray}\label{14}
&\overline{I^{E}_{{\rm tun},L}(t)}=\overline{\epsilon^{}_{d}(t)I^{}_{L}(t)}-\overline{I^{E}_{L}(t)}
\nonumber\\
&+2\Gamma_{L}\Gamma_{R}\int \frac{d\omega}{2\pi}\big(f_{R}(\omega)-f_{L}(\omega)\big) \Big[\overline{G_{dd}^{r}(\omega)}+\overline{G_{dd}^{a}(\omega)}\Big]\ .
\label{IRtun4}
\end{eqnarray}
For the two types of average, presented in Secs. \ref{TN} and \ref{EE}, this combination of averages vanishes.



\end{document}